\documentclass[prb,twocolumn,showpacs,floatfix]{revtex4}
\usepackage{graphicx,amsmath}

\begin{document}
\title{Large Photonic Band Gaps in Certain Periodic and Quasi-Periodic Networks in two and three dimensions}

\author{S.K.~Cheung, T.L.~Chan, Z.Q.~Zhang and C.T.~Chan}

\address{Department of Physics, Hong Kong University of Science and Technology, Clear Water Bay, Kowloon, Hong Kong}

\date{\today}

\begin{abstract}
The photonic band structures in certain two- and three-dimensional periodic networks made of one-dimensional waveguides are studied by using the
Floquet-Bloch theorem. We find that photonic band gaps exist only in those structures where the fundamental loop exhibits anti-resonant
transmission. This is also true for quasi-periodic networks in two and three dimensions, where the photonic band structures are calculated from
the spectra of total transmission arising from a source inside the samples. In all the cases we have studied, it is also found that the gap
positions in a network are dictated by the frequencies at which the anti-resonance occurs.
\end{abstract}


\maketitle
\begin{section}{Introduction}
In the last decade, significant effort has been devoted to the study of photonic band gap (PBG) materials, in which the dielectric constant is
periodic in space \cite{Joan95,Sokoulis01}. The existence of a spectral gap in which electromagnetic waves cannot propagate in any direction
offers the possibility to confine and control the propagation of electromagnetic waves. It can give rise to many interesting physical phenomena,
as well as wide applications in several scientific and technical areas. For examples, on the scientific side, it has been suggested that the
Anderson localization, which is complicated by the e-e interactions in an electronic system, is much easier to achieve inside the PBGs of a
disordered photonic system \cite{John87}. In practical applications, ultra-low-threshold semiconductor lasers lie in the existence of robust
microcavity modes in PBG systems, facilitated by the suppression of spontaneous emission of light \cite{Nature}.  Besides periodic structures,
PBG has also been found in certain quasi-periodic systems in two dimensions \cite{YSChan98,Cheng99,Zoorob00,XDZhang01}.

Recently, a different kind of band gap structure for EM wave systems has been proposed. These are systems of networks made of one-dimensional
waveguides \cite{Zhang94,Zhang98}, which are experimentally realizable, even in an undergraduate laboratory. In particular, by using coaxial
cable as the 1D waveguide, a large photonic band gap has been found in a 3D network in the diamond structure, in which certain 1D waveguides
have been replaced by resonant loops to produce strong scattering \cite{Zhang98}. In the presence of defect and randomness, both defect states
and the Anderson localized states have also been observed inside the gap.  Furthermore, network systems are also interesting for the studies of
many other potential physical problems such as PBG in fractals \cite{Li00}, and the wave localization in dimensions larger than three.

In this work, we have calculated the band structures of some basic lattice structures in two and three dimensions.  We have also studied the
photonic band structures in certain quasi-periodic networks in two and three dimensions by calculating the spectra of total transmission arising
from a source inside the samples. In all the cases we have studied, it is found that the existence of large gaps in a network relies mainly on
the presence of triangular loops in the fundamental building blocks of the lattice.  Each loop is capable of producing anti-resonances at
certain frequencies with zero transmission. These frequencies coincide with the positions of the gaps found in the network. The outline of this
paper is as follows. Section II describes the methods to calculate the band structures and the spectra of total transmission. The results of
certain periodic are also presented. Section III presents the results of certain quasi-periodic networks. Section IV gives the summary and
discussions.
\end{section}

\begin{section}{PBGs in Periodic Networks}
A network consists of certain nodes connected by segments of 1D waveguides.  The wavefunction at each node satisfies the following network
equation \cite{Zhang94}:

\begin{equation}
-\varphi_i\sum_j[\cot(\omega\ell_{ij}/c)]+\sum_j\left[\frac{1}{\sin(\omega\ell_{ij}/c)}\right]\varphi_j=0,
\end{equation}
where $\varphi_i$  is the scalar wavefunction at node $i$,  is the length of the 1D waveguide between nodes $i$ and $j$, $\omega$ and $c$ are,
respectively, the frequency and speed of the waves. The summation of $j$ is over all the nodes linked to node $i$. In the case of coaxial cable,
the function $\varphi_i$ represents the voltage at node $i$. If all the 1D waveguides have the same length, $\ell$ , it is easy to see from Eq.
(1) the network is periodic in $\omega$  with a period $2\pi c/\ell$. Eq. (1) is equivalent to a tight-binding Hamiltonian in an electronic
system with a zero energy eigenvalue, i.e.,

\begin{equation}
-\xi_i\varphi_i+\sum_jt_{ij}\varphi_j=0,
\end{equation}
where $\xi_i=\sum_j\cot k\ell_{ij}$  and $t_{ij}=1/\sin k\ell_{ij}$ are respectively, the correlated-site energy and the hopping energy.

\begin{figure}
\includegraphics [width=\columnwidth] {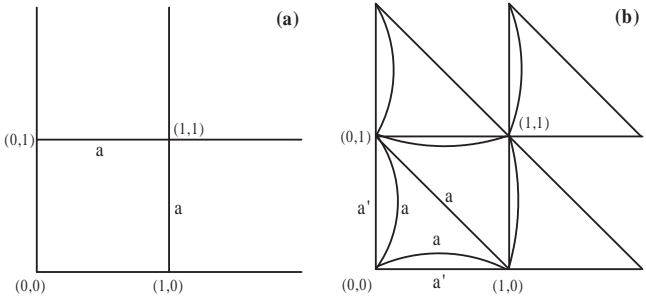}
\caption{The unit cell for a (a) square network, (b) triangular network. All segments in the unit cells have the same length $a$. The lattice
constants in (a) and (b) are respectively, $a$ and  $a'$.}
\end{figure}

\begin{figure}
\includegraphics [angle=90,width=\columnwidth] {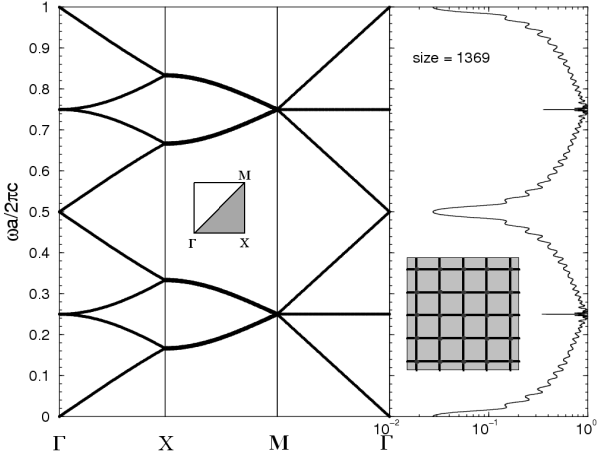}
\caption{Left panel: the band structure of the square network. Right panel: the total transmission of the square network with 1369 nodes. The
shaded region of the inset on the left panel shows the Brillouin zone.}
\end{figure}

The band structures or dispersion relations $\omega(\mathbf{\kappa})$ of any periodic network can be obtained by solving a set of coupled
equations in a unit cell with the use of the Floquet-Bloch theorem. As an example, we consider an infinite network with its nodes arranged in a
square lattice. Only nearest-neighbor nodes are connected by 1D waveguide of length $a$. Fig. 1a shows a unit cell with four nodes labeled by
(0,0), (1,0), (0,1) and (1,1). The period of the unit cell is chosen to be $2a$ for convenience, which is twice as large as a conventional unit
cell for a square lattice. The choice of a larger unit cell leads to a band folding effect and will not affect the structure of photonic band
gaps in the system. From Eq. (1), we can write down a set of networks equations for this unit as

\begin{eqnarray}
-4(\cos ka)\varphi_{(0,0)}+\varphi_{(0,-1)}+\varphi_{(0,1)}+\varphi_{(1,0)}+\varphi_{(-1,0)}&=&0\nonumber\\
-4(\cos ka)\varphi_{(0,1)}+\varphi_{(-1,1)}+\varphi_{(0,2)}+\varphi_{(0,0)}+\varphi_{(1,1)}&=&0\nonumber\\
-4(\cos ka)\varphi_{(1,0)}+\varphi_{(1,-1)}+\varphi_{(1,1)}+\varphi_{(0,0)}+\varphi_{(2,0)}&=&0\nonumber\\
-4(\cos ka)\varphi_{(1,1)}+\varphi_{(2,1)}+\varphi_{(1,2)}+\varphi_{(1,0)}+\varphi_{(0,1)}&=&0.\nonumber\\
\end{eqnarray}
Because of the discrete translational symmetry in a periodic network, by using the  Floquet-Bloch theorem, the wavefunctions at each node can be
written as

\begin{equation}
\varphi_{\kappa_x,\kappa_y}(m,n)=e^{i\kappa_xma}e^{i\kappa_yna}U(m,n),
\end{equation}
where $\kappa_x$ and $\kappa_y$ are the reciprocal lattice vectors or the Bloch wave vectors in $x$ and $y$-directions, respectively. The
indices $m$ and $n$ represent, respectively, the node indices in the $x$ and $y$-directions. $U(m,n)$ is a periodic function such that

\begin{equation}
U_{(m+2,n+2)}=U_{(m+2,n)}=U_{(m,n+2)}=U_{(m,n)}.
\end{equation}
By using Eqs. (4) and (5), Eq (3) becomes

\begin{equation}
\begin{pmatrix}
-2\cos ka & \cos \kappa_xa & \cos \kappa_ya & 0 \\
\cos \kappa_xa & -2\cos ka & 0 & \cos \kappa_ya \\
\cos \kappa_ya & 0 & -2cos ka & \cos \kappa_xa \\
0 & \cos \kappa_ya & cos  \kappa_xa & -2cos ka
\end{pmatrix}
\begin{pmatrix}
U_{(0,0)}\\U_{(0,1)}\\U_{(1,0)}\\U_{(1,1)}
\end{pmatrix}=0,
\end{equation}
where $k=\omega/c$ . For each Bloch wave vector $\mathbf{\kappa}=(\kappa_x,\kappa_y)$, the corresponding eigenfrequencies
$\omega(\mathbf{\kappa})$ are determined by the condition that the determinant in Eq. (6) vanishes, from which the band structure can be
obtained.

\begin{figure}
\includegraphics [angle=90,width=\columnwidth] {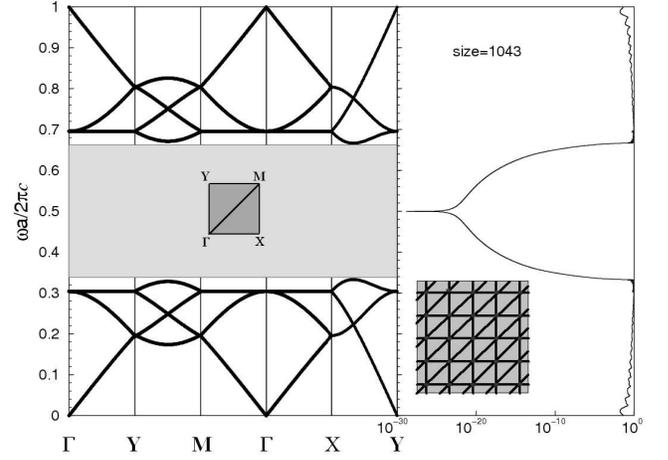}
\caption{Left panel: the band structure of the triangular network. Right panel: the total transmission of the triangular network with 1043
nodes. The shaded region of the inset on the left panel shows the Brillouin zone.}
\end{figure}

The band structure of the square network is shown in the left panel of Fig. 2, where the symmetry points $\Gamma$, $X$ and $M$ correspond to the
wave vectors $(\kappa_x, \kappa_y)$ = $(0,0)$, $(\pi/2a, 0)$, and $(\pi/2a, \pi/2a)$, respectively. This figure shows that the square network
does not possess a photonic band gap. It also shows that the band structures are periodic in $\omega$  with a period $\pi c/a$, which is half of
the period shown in Eq. (1). This reason is given in the following. A square lattice is bipartite, i.e., it has two interpenetrating
sublattices.  For each eigenfrequency of a given $\mathbf{\kappa}=(\kappa_x,\kappa_y)$, we can always obtain another eigenstate with frequency
$\omega+\pi c/a$ by changing the sign of all wavefunctions in one of the two sublattices. This can be seen easily from Eq. (6).  If we make the
following changes $ka\rightarrow ka+\pi$, $U(0,0)\rightarrow-U(0,0)$ and $U(1,1)\rightarrow-U(1,1)$, Eq. (6) remains unchanged. Such a reduction
in the period does not apply to the triangular network as it is not bipartite.

\begin{figure}
\includegraphics [width=\columnwidth] {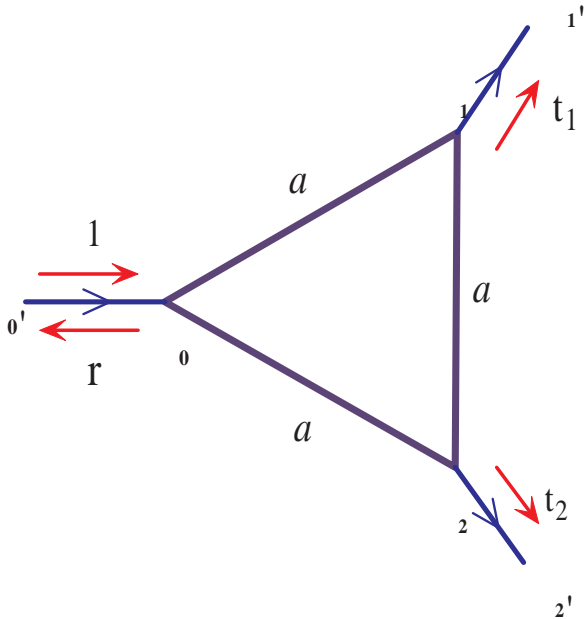}
\caption{A triangular loop with identical segment lengths $a$. Leads are connected at the input and output nodes.}
\end{figure}

To study the PBG structure in a triangular lattice, for convenience, we choose a unit cell that is plotted in Fig. 1b, where one additional
next-nearest-neighbor segment is inserted in each plaquette of a square lattice. However, this additional segment has the same length $a$, as
the nearest-neighbor ones. Thus the segments on the four sides of the square have to be bended and its lattice constant becomes $a'=a/\sqrt{2}$.
The unit cell shown in Fig. 1b produces a lattice that is topologically equivalent to the triangular lattice. The symmetry points $\Gamma$, $X$,
$Y$ and $M$ correspond to the wave vectors $(\kappa_x,\kappa_y)=(0,0)$, $(\pi/2a',0)$, $(0,\pi/2a')$ and $(\pi/2a',\pi/2a')$, respectively.
Similar to the case of square lattice, we can use Eq. (4) to write down a set of equations for the dispersion relations. The corresponding
secular equation now takes the form

\begin{equation}
\begin{vmatrix}
-3\cos ka & \cos \kappa_ya' & \cos \overline{\kappa}a' & \cos \kappa'a'\\
\cos \kappa_ya' & -3\cos ka & \cos \overline{\kappa}a' & \cos \kappa_xa'\\
\cos \kappa_xa' & \cos \overline{\kappa}a' & -3\cos ka & \cos \kappa_ya'\\
\cos \overline{\kappa}a' & \cos \kappa_xa' & \cos \kappa_ya' & -3\cos ka'
\end{vmatrix}=0,
\end{equation}
where $\overline{\kappa}=\kappa_x-\kappa_y$. The band structure calculated from Eq. (7) is shown in the left panel of Fig. 3.  A large PBG is
found with a mid-gap frequency at $\omega_c=\pi c/a$. Unlike the case of square network, the period of this band structure is $2\pi c/a$  as
dictated by Eq. (1) when all segments in a network have the same length $a$.

The photonic band gap in a network system can also be studied by calculating the spectra of total transmission arising from a source inside the
samples. This method is particularly useful for searching PBG when the structure has no translational symmetry (e.g. quasi-periodic or fractal
networks, etc.). In this method, we inject wave energy from a particular node inside a finite-sized network and connect all boundary nodes by
leads, which allow waves to leak out. For propagating states, the injected waves can propagate to the boundary nodes and give a finite total
transmission. However, if the frequency is inside a PBG, the injected waves cannot propagate to the boundary node and, therefore, give a
vanishing total transmission. Thus, a gap in the spectrum of total transmission is a signature of a PBG. Below we use a simple network to show
how to calculate the total transmission arising from a given source.

Consider an equilateral triangular loop connected by a lead at each node shown in Fig. 4. The network equations of this system have the
following forms:

\begin{eqnarray}
\varphi_{0'}-(3\cos ka)\varphi_0+\varphi_1+\varphi_2&=&0\nonumber\\
\varphi_{1'}-(3\cos ka)\varphi_1+\varphi_2+\varphi_0&=&0\nonumber\\
\varphi_{2'}-(3\cos ka)\varphi_2+\varphi_0+\varphi_1&=&0,
\end{eqnarray}
where $a$ is the length of each segment and $\varphi_{i'}$ is the wavefunction in the lead at a distance $a$ away from the Node $i$. If the
waves are injected from Node $0$ and transmitted through Nodes $1$ and $2$, the wavefunction in the lead at Node $0$ can be expressed as
$\exp(iks_0)+r\exp(-iks_0)$, where $s_0$ has the origin at Node $0$. Similarly, the wavefunctions in the leads at Nodes $1$ and $2$ can be
expressed as $t_1\exp(iks_1)$ and $t_2\exp(iks_2)$, respectively. Here $r$, $t_1$ and $t_2$ are respectively, the reflection and transmission
amplitudes. Using these expressions, we have $\varphi_{0'}=\exp(-ika)+r\exp(ika)$, $\varphi_{1'}=t_1\exp(ika)$ and $\varphi_{2'}.=t_2\exp(ika)$.
By substituting the above expressions into Eq. (8), we solve the transmission problem of the network, from which we find the reflection and
transmission coefficients as $R=|r|^2$, $T_1=|t_1|^2$ and $T_2=|t_2|^2$. The flux conservation requires that $R+T_1+T_2=1$. It should be noted
that the reflection coefficient $R$ and the total transmission $T=T_1+T_2$ will not depend on the choice of positions $0'$, $1'$ and $2'$ in the
leads, i.e., these positions can be arbitrary. Now we apply the above method to a triangular network containing 1043 nodes. The wave is injected
from a node located at the center of the sample. Transmission leads are connected at each node on the boundary. The spectrum of total
transmission is shown in the right panel of Fig. 3. A large gap is found in the spectrum of total transmission, which coincides with that of
band structure calculation. Thus, both the band structure and total transmission show a large band gap with a mid-gap frequency at $\omega_c=\pi
c/a$ and a width of $\Delta\omega=0.64\omega_c$. The spectrum of total transmission for the square network is shown in the right panels of Fig.
2. No gap is found in this structure. This is also consistent with the band structure calculation. It should be noted that there do exist dips
in the spectrum, but the total transmission at the dip is not exponentially small compared to the gap in the triangular network in Fig. 3. The
dip is resulting from the small density of states around the corresponding frequency.

\begin{figure}
\includegraphics [angle=90, width=\columnwidth] {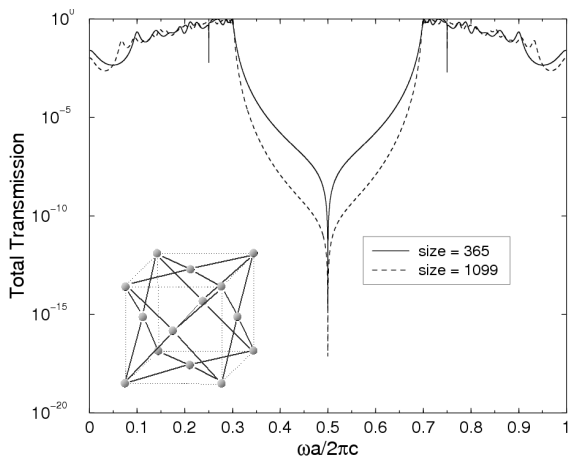}
\caption{The total transmissions of the fcc networks with 365 (solid line) and 1099 nodes (dash line).}
\end{figure}

In general, it is more difficult to find large full gaps in three-dimensions because its existence requires the pseudo-gaps in each direction to
overlap with a large extent. For the network system, PBG can still exist in certain networks in three dimensions. The reason will be given later
in Section IV. Since the band structure of 3D networks is rather complicated, here we present the total transmission instead. The total
transmissions of a fcc network shown in Fig. 5 indicates that there is a large band gap with a mid-gap frequency $\omega_c=\pi c/a$. The width
of this band gap is $\Delta\omega=0.8\omega_c$ , which is 25\% larger than that of the triangular network. In Fig. 5, the total transmissions of
two sample sizes are plotted, where the solid line and dash line are, respectively, the total transmission of the fcc networks with 365 nodes
and 1099 nodes.  It is seen that gap position remains unchanged when the number of nodes is increased 3 times. However, the total transmission
inside the gap of the larger sample is lower as expected. In addition to the periodic structures presented above, we have also studied the
honeycomb network in 2D, the simple cubic, bcc and diamond networks in 3D. However, no band gap is found in these structures.
\end{section}

\begin{figure}
\includegraphics [angle=270, width=\columnwidth] {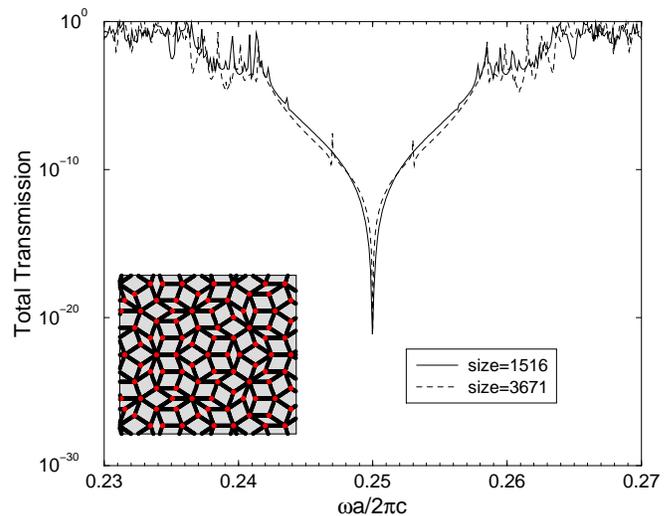}
\caption{The total transmissions of the pentagonal networks with 1517 (solid line) and 3671 nodes (dash line).}
\end{figure}

\begin{section}{PBGs in Quasi-Periodic Networks}
It is well known that quasi-periodic structures do not have spatial periodicity, and thus photonic band gaps that appear in quasi-periodic
structure may not have the same physical mechanism as in the case of periodic lattice by Bragg diffraction. However, several references have
reported PBGs in quasi-periodic structures \cite{YSChan98,Cheng99,Zoorob00,XDZhang01}. It was also suggested that a perfect long range periodic
order may not be necessary to create a spectral gap in a dielectric PBG system provided that strong Mie-resonance scattering takes place at
scattering centers.  In the network system, we found that the existence of photonic band gap is quite insensitive to both the periodicity and
dimensionality of the lattice, but is strongly correlated with the Mie-like scattering inside a single loop of waveguides. To illustrate this
point, we have studied three kinds of quasi-periodic networks. The two of them are two-dimensional networks and are respectively a pentagonal
structure formed by Penrose tiling, with a five-fold symmetry \cite{Levine86,Socolar86}, and a dodecagonal structure formed by square-triangle
tiling, with a twelve-fold symmetry \cite{Oxborrow93}. The third case is the three-dimensional network with icosahedral structure consisting of
icosahedral symmetry that contains 15 two-fold, 10 three-fold and 6 five-fold rotational axes. In this study, the icosahedral lattice is
constructed by the six-dimensional hypercube projection method \cite{Mihalkovic96}. The constructed structure has a node at the center of an
icosahedron. The distance between the nodes on the faces and the node at the center of the icosahedron is given by

\begin{eqnarray}
\eta=\frac{\sqrt[4]{5}\sqrt{\tau}}{2}a=0.9511\nonumber,
\end{eqnarray}
where $a$ is the distance between two nearby nodes at the faces and $\tau$ is the golden mean. For convenience, the nodes at the faces and the
node at the center will also be considered as the nearest neighboring pairs, so that the largest coordination number of the icosahedral network
is 12. We first consider the two-dimensional structures. The total transmissions of the pentagonal quasi-periodic networks, with 1516 (solid
line) and 3671 nodes (dash line) are shown in Fig. 6. It is interesting to see that a large gap exists when sample size is 1516 nodes. However,
when the sample size is increased to 3671 nodes, new states appear inside the gap, which are characterized by the appearance of local peaks in
the total transmission. These states are found to be localized states. For instance, the wave intensity profile of a localized state at the
normalized frequency $\omega a/2\pi c=0.247$ is plotted in Fig. 7. Due to the appearance of localized states at large samples, it is possible
that this gap would be continuously diminished when the size of the network is further increased. A different behavior is found in the
dodecagonal quasi-periodic network. The total transmissions of the dodecagonal networks with 289 (solid line) and 4123 nodes (dash line) are
shown in Fig. 8. A large photonic band gap exists at $\omega_c=\pi c/a$. The width of this band gap is $\Delta\omega=0.6\omega_c$. This gap
remains robust when the size of the network is increased from 289 to 4123 nodes. This suggests that a bona-fide large band gap can exist in the
dodecagonal network.

\begin{figure}
\includegraphics [angle=90,width=\columnwidth] {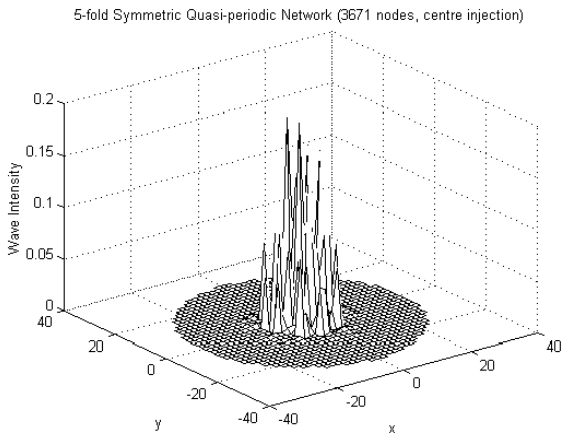}
\caption{The localized intensity distribution at frequency $\omega=0.247(2\pi c/a)$, for the pentagonal network with 3671 nodes.}
\end{figure}

\begin{figure}
\includegraphics [angle=270, width=\columnwidth] {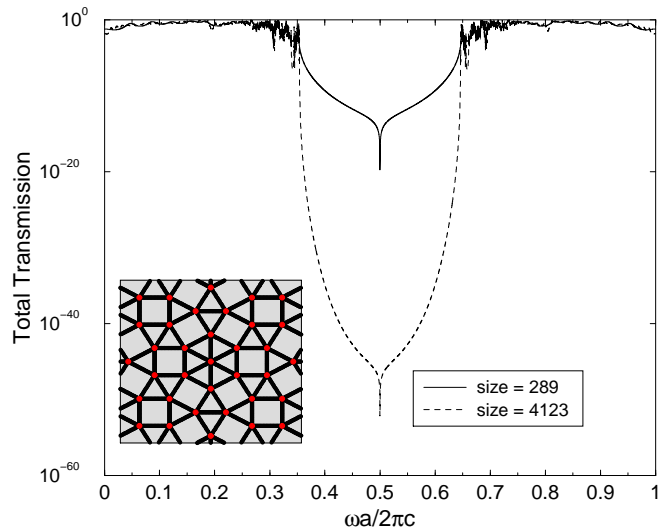}
\caption{The total transmissions of the dodecagonal networks with 289 (solid line) and 4123 nodes (dash line).}
\end{figure}

In three-dimensions, the icosaherdral quasi-periodic network is also found to posses a large photonic band gap at the mid-gap frequency
$\omega_c=\pi c/a$, according to the total transmission shown in Fig. 9. The number of nodes of the icosahedral networks are 931 (solid line)
and 2675 (dash lines). The width of the band gap is $\Delta\omega=0.78\omega_c$ , which is 30\% larger than that of the dodecagonal
quasi-periodic network, but is slightly smaller than that of the fcc network.
\end{section}

\begin{figure}
\includegraphics [angle=90, width=\columnwidth] {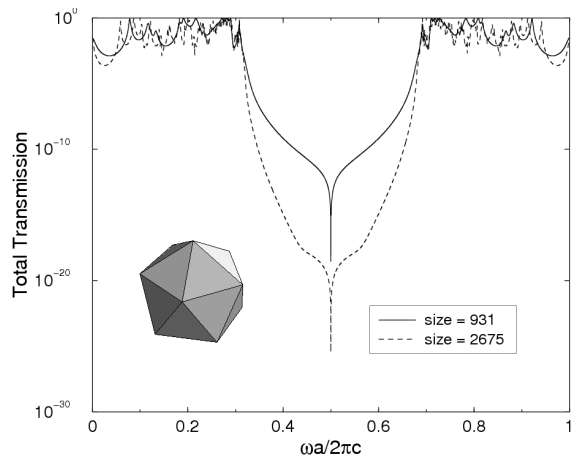}
\caption{The total transmissions of the icosaherdral quasi-periodic networks with 931 (solid line) and 2675 nodes (dash line).}
\end{figure}

\begin{table*}
\caption{The summary of photonic band gaps found in various networks. $\omega_c$ is the midgap frequency and $\Delta\omega$ is the width of the
band gap.}
\begin{ruledtabular}
\begin{tabular}{ccccc}
Networks & Dimensions & Periodicity & Fundamental building blocks & Band gap $\Delta/\omega_c$\\
\hline
Square & 2 & Periodic & Rhombus & Nil\\
Triangular & 2 & Periodic & Triangle & 0.64\\
Honeycomb & 2 & Periodic & Hexagon & Nil\\
Diamond & 3 & Periodic & Hexagon & Nil\\
Simple cubic & 3 & Periodic & Rhombus & Nil\\
bcc & 3 & Periodic & Rhombus & Nil\\
fcc & 3 & Periodic & Triangle, rhombus & 0.8\\
Pentagonal & 2 & Quasi-periodic & Rhombus & Nil\footnotemark[1]\\
Dodecagonal & 2 & Quasi-periodic & Triangle, Rhombus & 0.6\\
Icosahedral & 3 & Quasi-periodic & Triangle, Rhombus & 0.78
\end{tabular}
\end{ruledtabular}
\footnotetext[1]{PBG is absence because localized states appear inside the band gap when the size is large.}
\end{table*}

\begin{figure}
\includegraphics [angle=270, width=\columnwidth] {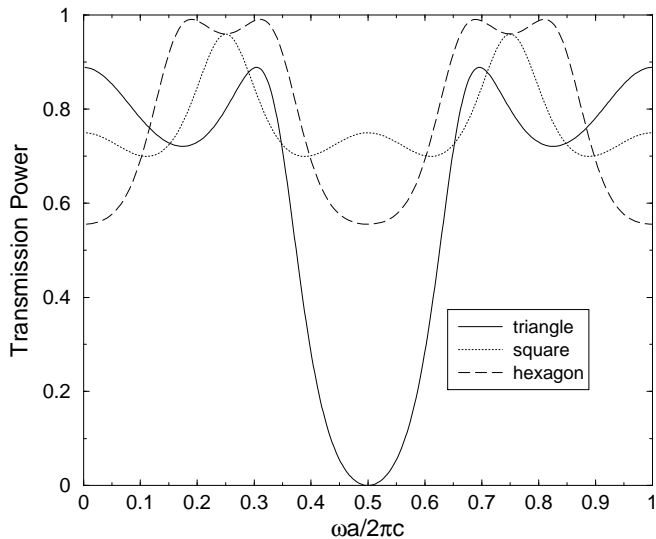}
\caption{The transmission spectra of the triangle (solid line), square (dotted line) and hexagon (dash line) loops, which are respectively, the
fundamental units in triangular, square and honeycomb networks.}
\end{figure}

\begin{section}{Discussion and Conclusions}
It is remarkable that photonic band gaps exist in a number of network topologies without introducing any resonant loops. We summarize the
intrinsic band gap properties of various periodic and quasi-periodic networks in Table 1, where the fundamental building blocks and size/mid-gap
ratio are listed and compared. In this Table, it is obvious that only the networks that contain triangular loops in the fundamental building
block possess photonic band gaps.  This suggests that the existence of the large PBGs is related to the wave interference effect inside
triangular loops in close analogy to the Mie scattering by spherical scattering centers \cite{Lidorikis98}. To confirm this observation, the
transmission spectra of three loops with different connectivity are shown in Fig. 10. They are the triangular (solid line), square (dash line)
and hexagonal loops (dotted line), which are the fundamental units in triangular, square and honeycomb networks. The transmission spectra are
calculated by injecting waves to one of the leads and summing up the transmission coefficients at all others.  From Fig. 10, it is found that
although resonances appear in the square and hexagon loops, however, these resonances are not strong enough to produce a gap in any frequency in
the respective networks. In addition to resonances, the triangular loop is capable of producing an anti-resonance at frequency  $\omega_c=\pi
c/a$, where the total transmission is zero. This zero-transmission frequency is precisely the mid-gap frequency found in all the structures that
possess gap in Table 1. The full-width-half-maximum of the transmission dip in triangular loop is $0.5\omega_c$.  This is also roughly the size
of the gaps shown in Table. It should be pointed out that anti-resonance can exist only in ring-shape loops with odd number of nodes. The reason
is as follows. For an even-node loop, an incident wave can propagate along two identical paths and interfere constructively at the node in the
opposite side of the loop, thus, giving non-vanishing contribution to the total transmission.

In summary, we have investigated the propagation of waves in several network structures in 2D and 3D, including both periodic and quasi-periodic
structures. Certain networks have large robust photonic band gap, namely the triangular, fcc, dodecagonal (12-fold) and icosahedral networks.
The existence of such PBGs is found to be related to the existence of an anti-resonance inside triangular loops of the fundamental building
units. The gap positions in a network are also dictated by the frequencies at which the anti-resonant transmission occurs. What we have studied
here are intrinsic PBGs in network systems that do not have additional resonant loops.  For those systems that do not have a gap, one can always
create a gap by introducing additional resonant loops into the system as shown in Ref. \cite{Zhang98}.  The gap size as well as its position can
also be 'tailor-made' by adjusting the resonant structure of the resonant loop.
\end{section}

\begin{acknowledgments}
This work was supported by Hong Kong RGC Grant No.6112/98P.
\end{acknowledgments}

\end{document}